\documentclass{article}
\usepackage{spconf,amsmath,graphicx}
\usepackage{setspace}
\usepackage{breqn}
\usepackage[utf8]{inputenc}
\usepackage{amsmath}
\usepackage{amssymb}
\usepackage{xspace}
\usepackage{subfigure}
\usepackage{algorithm}
\usepackage{algpseudocode}
\usepackage{tabularx}
\usepackage{multirow}
\usepackage{booktabs}
\usepackage{hyperref}
\usepackage{footmisc}

\usepackage{etoolbox}
\patchcmd{\thebibliography}
  {\settowidth}
  {\setlength{\itemsep}{0pt plus 0.1pt}\settowidth}
  {}{}
\apptocmd{\thebibliography}
  {\small}
  {}{}

\usepackage{xcolor}

\definecolor{YenjuColor}{rgb}{1.0,0.0,0.0}
\definecolor{ShinjiColor}{rgb}{0.0,0.0,0.8}
\definecolor{AlexColor}{rgb}{0.8,0.8,0.0}

\usepackage{fancyhdr}
\makeatletter
\DeclareRobustCommand\onedot{\futurelet\@let@token\@onedot}
\def\@onedot{\ifx\@let@token.\else.\null\fi\xspace}

\def\ie{\emph{i.e}\onedot}

\makeatother

\title{Conditional Diffusion Probabilistic Model for Speech Enhancement}
\name{Yen-Ju Lu$^{1,3}$, Zhong-Qiu Wang$^{1}$, Shinji Watanabe$^{1}$, Alexander Richard$^{2}$, Cheng Yu$^{3}$, and Yu Tsao$^{3}$}

\address{
  $^1$Language Technology Institute, Carnegie Mellon University, Pittsburgh, PA, USA\\
  $^2$Reality Labs Research, Pittsburgh PA, USA\\
  $^3$Research Center for Information Technology Innovation, Academia Sinica, Taipei, Taiwan \\
  }
\begin{document}
\ninept
\thispagestyle{fancy}
\maketitle
\setlength{\abovedisplayskip}{2.5pt}
\setlength{\belowdisplayskip}{2.5pt}
\begin{abstract}
Speech enhancement is a critical component of many user-oriented audio applications, yet current systems still suffer from distorted and unnatural outputs.
While generative models have shown strong potential in speech synthesis, they are still lagging behind in speech enhancement. This work leverages recent advances in diffusion probabilistic models, and proposes a novel speech enhancement algorithm that incorporates characteristics of the observed noisy speech signal into the diffusion and reverse processes. 
More specifically, we propose a generalized formulation of the diffusion probabilistic model named conditional diffusion probabilistic model that, in its reverse process, can adapt to non-Gaussian real noises in the estimated speech signal. In our experiments, we demonstrate strong performance of the proposed approach compared to representative generative models, and investigate the generalization capability of our models to other datasets with noise characteristics unseen during training.



\end{abstract}


\section{Introduction}
\label{sec:intro}

Speech enhancement, a key element for 
immersive audio experiences in telecommunication as well as a crucial front-end processor for
robust speech recognition~\cite{li2014overview, HaebUmbach2020Overview}, assistive hearing ~\cite{healy2019optimal},
and robust speaker recognition~\cite{H.L.Hansen2015, michelsanti2017conditional}, is a challenging and still unsolved problem in audio processing.
%
%
Riding on the advance of deep learning, considerable progress has been made in the past decade~\cite{wang2018supervised,lu2013speech}.
%
%
Deep learning based approaches can be roughly divided into two categories, based on the criteria used to estimate the transformation function from noisy-reverberant speech to clean speech.
The first category trains discriminative models to minimize the difference between enhanced and clean speech, where the difference can be
a point-wise $L_p$-norm distance \cite{fu2018end},
or can be computed based on a perceptual metric~\cite{koizumi2018dnn, fu2019metricgan}.
The second category considers the distribution of the clean speech signals to form the objective function.
Well-known examples along this direction include generative adversarial networks (GANs)~\cite{pascual2017segan,soni2018time}, Bayesian wavenet~\cite{qian2017speech}, variational autoencoders~\cite{leglaive2020recurrent}, and flow-based models~\cite{strauss2021flow}.
While the best performing approaches typically fall into the first category~\cite{defossez2020real,wang2018supervised},
they usually introduce unpleasant speech distortion and phonetic inaccuracies to the enhanced speech~\cite{wang2019bridging,bagchi2018spectral,gao2016snr}.
%
Generative approaches that aim to match the distribution of speech signals
rather than regressive approaches optimizing a point-wise loss hold the promise to produce more natural sounding speech, although they are currently lagging behind regressive approaches and require more research to unfold their potential.

This work investigates diffusion probabilistic models~\cite{sohl2015deep}, a class of generative models that have shown outstanding performance in image generation~\cite{ho2020denoising,nichol2021improved} and audio synthesis~\cite{kong2020diffwave,liu2021diffsvc,lu2021study}, for speech enhancement.
Diffusion probabilistic models convert clean input data to an isotropic Gaussian distribution in a step-by-step diffusion process and, in a reverse process, gradually restore the clean input by predicting and removing the noise introduced in each step of the diffusion process.
%
%
These models, in their vanilla formulation, assume isotropic Gaussian noise in each step of the diffusion process as well as the reverse process.
%
However, in realistic conditions,
the noise characteristics
are usually non-Gaussian,
which violates the model assumption when directly combining the noisy speech signal in the sampling process.
We address this problem by formulating a generalized conditional diffusion probabilistic model that incorporates the observed noisy data into the model.
We derive the corresponding conditional diffusion and reverse processes as well as the evidence lower bound (ELBO) optimization criterion \cite{ho2020denoising}, and show that the resulting model is a generalization of the original diffusion probabilistic model.
In our experiment, we will demonstrate that our formulation can not only improve over the vanilla diffusion probabilistic model, but also outperform other generative models.
%

\section{Diffusion Probabilistic Model}

A $T$-step diffusion model~\cite{ho2020denoising} consists of two processes: the \textit{diffusion} process with steps $t \in \{0,1,\cdots,T\}$ and the \textit{reverse} process $t \in \{T,T-1,\cdots,0\}$.
We start with a brief summary of the vanilla diffusion probabilistic model, \ie, we revisit the original diffusion- and reverse process.

\subsection{Diffusion Process}
\label{subsec:diffusion_process}

Given the clean speech data $x_0$, the diffusion process $q_{\text{data}}(x_0)$ of the first diffusion step ($t=0$) is defined as the data distribution $x_0$ on $\mathbb{R}^L$, where $L$ is the signal length in samples. For the the $t$-th diffusion step, we have a step-dependent variable $x_t\in \mathbb{R}^L$ with the same signal length $L$. 
The diffusion process from data $x_0$ to the variable $x_T$ can be formulated based on a fixed Markov chain:
\begin{align}
  q(x_1,\cdots,x_T|x_0) 
  & = \prod_{t=1}^{T} q(x_t|x_{t-1}),
  \label{diffuse eq1}
\end{align}
with a Gaussian model $q(x_t|x_{t-1}) = \mathcal{N}(x_t;$ $\sqrt{1-\beta_{t}}x_{t-1},\beta_{t}I)$, where $\beta_{t}$ is a small positive constant.
In other words, in each step a Gaussian noise is added to the previous sample $x_{t-1}$.
According to the pre-defined schedule $\beta_{1},\cdots,\beta_{T}$, the overall process gradually converts clean $x_0$ to a latent variable with an isotropic Gaussian distribution of $p_{\text{latent}}(x_T) = \mathcal{N}(0,I)$.

By substituting the Gaussian model of $q(x_t|x_{t-1})$ into Eq.~\eqref{diffuse eq1} and by marginalizing $x_{1}, \dots, x_{t-1}$, the sampling distribution of $ x_t$ can be derived as the following distribution conditioned on $x_0$:
\begin{equation}
  q(x_t|x_0) = \mathcal{N}(x_t;\sqrt{\bar{\alpha}_t}x_0,(1-\bar{\alpha}_t)I),
  \label{diffuse eq2}
\end{equation}
where $\alpha_t = 1-\beta_t$ and $\bar{\alpha}_t =  \prod_{s=1}^{t} \alpha_s$.

\subsection{Reverse Process}
\label{sec:reverse_process}

The reverse process converts the latent variable $x_{T} \sim N(0,I)$ to $x_{0}$, also based on a Markov chain similar to Eq.~\eqref{diffuse eq1}:
\begin{equation}
  p_\theta(x_0,\cdots,x_{T-1}|x_T) = \prod_{t=1}^{T} p_\theta(x_{t-1}|x_t),
  \label{reverse eq1}
 \end{equation}
where $p_\theta(\cdot)$ is the distribution of the reverse process with learnable parameters $\theta$.
Unlike the diffusion process, the following marginal likelihood is intractable:
\begin{align}
    p_{\theta}(x_0) = \int p_\theta(x_0, \cdots, x_{T-1}|x_T)\cdot p_{\text{latent}}(x_T) dx_{1:T}.
\end{align}
Therefore, we use the ELBO to form an approximated objective function for model training. In ~\cite{ho2020denoising}, it is reported that minimizing the following equation leads to higher generation quality:
\begin{equation}
    c + \sum_{t=1}^{T}\kappa_t\mathbb{E}_{x_0,\epsilon}  \parallel\epsilon - \epsilon_\theta(\sqrt{\bar{\alpha}_t}x_0 + \sqrt{1-\bar{\alpha}_t}\epsilon,t) \parallel^2_2,
  \label{eq:ELBO_optimized_original}
\end{equation}
with constants $c$ and $\kappa_t$. Here $\epsilon_\theta$ is the model trained to estimate the Gaussian noise $\epsilon$ in $x_t$.
After optimizing Eq.~\eqref{eq:ELBO_optimized_original}, the corresponding reverse process equation becomes:
\begin{equation}
 p_{\theta}(x_{t-1}|x_t) =
 \mathcal{N}(x_{t-1};\mu_\theta(x_t,t),\tilde{\beta}_t I),
 \label{eq:reverse_solution_original}
\end{equation}
where the mean $\mu_\theta(x_t,t)$ is: 
\begin{equation}
\mu_\theta(x_t,t)= \frac{1}{\sqrt{\alpha_t}}(x_t-\frac{\beta_t}{\sqrt{1-\bar{\alpha}_t}}\epsilon_\theta(x_t,t)). 
 \label{eq:reverse_mean_original}
\end{equation}
The $\mu_\theta(x_t,t)$ predicts the mean of $x_{t-1}$ distribution by removing the estimated Gaussian noise $\epsilon_\theta(x_t,t)$ in the $x_t$, and the variance is fixed to a constant $\tilde{\beta}_t = \frac{1-\bar{\alpha}_{t-1}}{1-\bar{\alpha}_t}\beta_t$. 

\section{Conditional Diffusion Probabilistic Model}

\begin{figure}[tp]
 \centering
 \includegraphics[width=7cm]{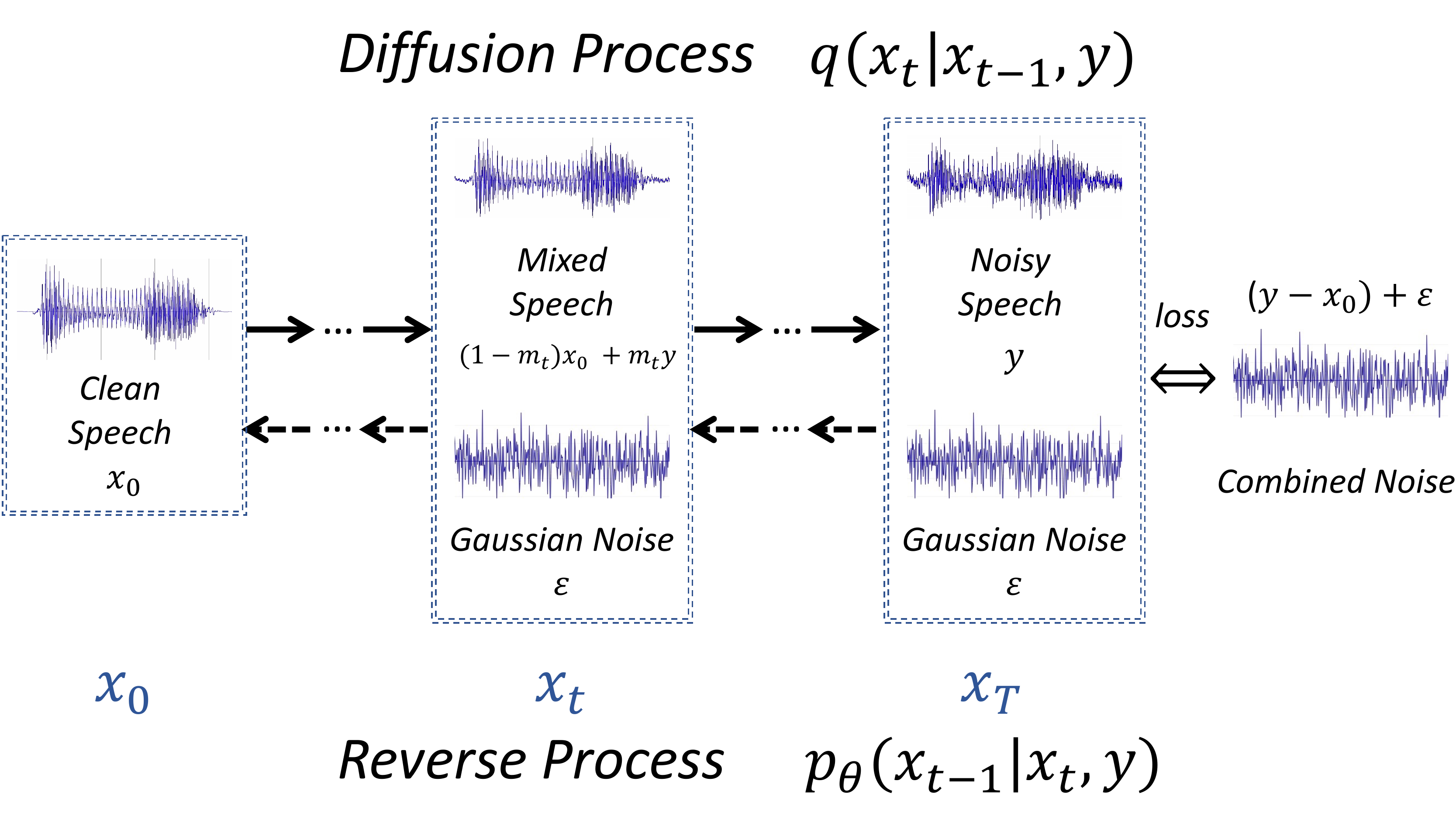} \vspace{-0.3cm}
 \caption{Diffusion process (solid arrows) and reverse process (dashed arrows) of the proposed conditional diffusion probabilistic model.}
 \label{fig:new diffusion model}
 \vspace{-0.3cm}
\end{figure}

The original diffusion process in Section \ref{subsec:diffusion_process} starts from the clean data $q_{\text{data}}(x_0)$ and adds Gaussian noise into the speech signal. In the proposed conditional diffusion probabilistic model, we incorporate the noisy data $y$ into the diffusion process, as shown in Fig.~\ref{fig:new diffusion model}.

\subsection{Conditional Diffusion Processes}
\label{Sec:CDiffuSE diffusion}

In the conditional diffusion process, we use an interpolation parameter $m_t$ to combine the clean data $x_0$ and the noisy data $y$, the summation of $x_0$ and the real noise $n$, on the solid arrows in Fig.~\ref{fig:new diffusion model}.
Instead of starting from the Markov chain Gaussian model $q(x_t|x_{t-1})$ in the original diffusion process, we first define the following conditional diffusion process $q(x_t|x_0,y)$: 
\begin{equation}
  q_{\text{cdiff}}(x_t|x_0,y) = \mathcal{N}(x_t;(1-m_t)\sqrt{\bar{\alpha}_t}x_0 + m_t\sqrt{\bar{\alpha}_t}y ,\delta_tI),
  \label{eq:new diffusion process t}
\end{equation}
where $\delta_t$ is the variance.
Unlike the original diffusion process $q(x_t|x_0)$ in Eq.~\eqref{diffuse eq2}, we assume that the Gaussian mean in Eq.~\eqref{eq:new diffusion process t} is represented as a linear interpolation between the clean data $x_0$ and the noisy data $y$ with the interpolation ratio $m_t$.
$m_t$ starts from $m_0 = 0$ and is gradually increased to $m_T \approx 1$, turning the mean of $x_t$ from the clean speech $x_0$ to noisy speech $y$  as in Fig.~\ref{fig:new diffusion model}.

%

Given the interpolation formulation in Eq.~\eqref{eq:new diffusion process t}, we can derive $q_{\text{cdiff}}(x_t|x_0)=\int q_{\text{cdiff}}(x_t|x_0,y) p_y(y|x_0) dy$ by marginalizing $y$ in the multiplication of $q_{\text{cdiff}}(x_t|x_0,y)$ and $p_y(y|x_0)$ with the special case where $n \sim \mathcal{N}(0,I)$. 
Then, $q_{\text{cdiff}}(x_t|x_0)$ becomes equivalent to the original diffusion process $q(x_t|x_0)$ in Eq. ~\eqref{diffuse eq2} when
\begin{equation}
    \delta_t = (1-\bar{\alpha}_t)-m_t^2\bar{\alpha}_t.
  \label{eq:new diffusion process t delta}
\end{equation}
This analytical result indicates that our model is a generalization of the original diffusion probabilistic model. 
In our previous study \cite{lu2021study}, we investigated directly utilizing noisy signal in the reverse process; 
the idea is found to work well empirically, but there lacks a theoretical justification. In Sec.~\ref{Sec:CDiffuSE reverse}, we will propose a conditional reverse process that is theoretically sound.
To further research the effect of incorporating noisy signal in the diffusion model, in Sec.~\ref{sec:elbo}, we will set $\delta_t$ according to Eq.~\eqref{eq:new diffusion process t delta} so that the conditional diffusion process becomes a generalized version of the original diffusion process\footnote{It is difficult to derive $q_{\text{cdiff}}(x_t|x_{t-1})$ for satisfying the original diffusion process if we first define $q_{\text{cdiff}}(x_t|x_{t-1},y)$, because the distribution of $y$ depends on $x_0$ as $y = x_0 + n$. \label{fn1}}.
\subsection{Conditional Reverse Processes}
\label{Sec:CDiffuSE reverse}

In the conditional reverse process, we start from $x_T$, noisy speech signal $y$ with variance $\delta_T$, according to Eq. \eqref{eq:new diffusion process t} with $m_T = 1$:
\begin{equation}
  p_{\text{cdiff}}(x_{T}|y) = \mathcal{N}(x_T,\sqrt{\bar{\alpha}_T}y,\delta_TI).
  \label{eq:new reverse xT}
\end{equation}
Based on the Markov chain, similar to Eq. \eqref{eq:reverse_solution_original}, the conditional reverse process on the dashed arrows in Figure \ref{fig:new diffusion model} aims to predict $x_{t-1}$ based on $x_{t}$ and $y$:
\begin{equation}
  p_{\text{cdiff}}(x_{t-1}|x_t,y) = \mathcal{N}(x_{t-1};\mu_\theta(x_t,y,t),\tilde{\delta}_tI),
  \label{eq:new reverse}
\end{equation}
where the $\mu_\theta(x_t,y,t)$ is the estimated mean of the conditional reverse process.
The concrete form of the variance $\tilde{\delta}_t$ is introduced later.
In contrast to the vanilla diffusion model, we further condition the diffusion model on $y$. Therefore, similar to Eq. \eqref{eq:reverse_mean_original}, the mean $\mu_\theta(x_t,y,t)$ in each reverse step is a linear combination of $x_t$, $y$, and estimated noise $\epsilon$ with weights $ c_{xt}, c_{yt} $ and $ c_{\epsilon t} $,
\begin{equation}
  \mu_\theta(x_t,y,t) = c_{xt}x_{t} + c_{yt} y - c_{\epsilon t} \epsilon_\theta(x_t,y,t),
  \label{eq:new reverse mean}
\end{equation}
where the $\epsilon_\theta(x_t,y,t)$ is the model to estimate the Gaussian and non-Gaussian noise combination.
The coefficients $ c_{xt}, c_{yt}, $ and $ c_{\epsilon t} $ can be derived from the ELBO optimization criterion, see Section~\ref{sec:elbo}.

\subsection{Coefficient Estimation by Optimizing ELBO}
\label{sec:elbo}


By modifying the derivations in \cite{ho2020denoising}, we obtain the ELBO condition for the conditional diffusion process to optimize the likelihood:
\begin{dmath}
 ELBO =-\mathbb{E}_q \Big(D_{\text{KL}}(q_{\text{cdiff}}(x_T|x_0,y)||p_{\text{latent}}(x_T|y)) \\
 + \sum_{t=2}^{T}D_{\text{KL}}(q_{\text{cdiff}}(x_{t-1}|x_t,x_0,y)||p_\theta(x_{t-1}|x_t,y)) \\
 - \log p_\theta(x_0|x_1,y) \Big).
  \label{eq:ELBO KL}
\end{dmath}
To optimize Eq.~\eqref{eq:ELBO KL}, we first need the distribution $q_{\text{cdiff}}(x_t|x_{t-1},y)$. 
Generally, the diffusion process define $q_{\text{cdiff}}(x_t|x_{t-1},y)$ first and derive $q_{\text{cdiff}}(x_t|x_0,y)$ by marginalizing $x_0, \cdots, x_{t-1}$ . Instead, we first design the interpolation form in Eq. \eqref{eq:new diffusion process t} 
as mentioned in Sec.~\ref{Sec:CDiffuSE diffusion}. 
%
Therefore, we compare the coefficients of the marginalized result and Eq. \eqref{eq:new diffusion process t} to compute the coefficients of $q_{\text{cdiff}}(x_t|x_{t-1},y)$ as:
\begin{align}
  & q_{\text{cdiff}}(x_t|x_{t-1},y) = \mathcal{N}\Bigl(x_t;\frac{1-m_t}{1-m_{t-1}}\sqrt{\alpha}_t x_{t-1} \nonumber \\
  & \quad \quad \quad \quad + \Bigl(m_t - \frac{1-m_t}{1-m_{t-1}}m_{t-1}\Bigr)\sqrt{\bar{\alpha}}_t y ,\delta_{t|t-1}I\Bigr),
  \label{eq:new diffusion process t, t-1}
\end{align}
where the $\delta_{t|t-1}$ is also calculated by $\delta_t$ to satisfy Eq. \eqref{eq:new diffusion process t delta} as:
\begin{equation}
    \delta_{t|t-1} = \delta_t - \left(\frac{1-m_t}{1-m_{t-1}}\right)^2 \alpha_t \delta_{t-1}.
  \label{eq:new diffusion process mean t, t-1}
\end{equation}
%

%

Then, by combining Eqs. \eqref{eq:new diffusion process t} and \eqref{eq:new diffusion process t, t-1},  
$q_{\text{cdiff}}(x_{t-1}|x_t,x_0,y)$ can be derived through Bayes' theorem and the Markov chain property:
\begin{dmath}
  q_{\text{cdiff}}(x_{t-1}|x_t,x_0,y) = \mathcal{N}\Bigl(x_{t-1};\frac{1-m_t}{1-m_{t-1}}\frac{\delta_{t-1}}{\delta_{t}}\sqrt{\alpha}_t x_{t} + (1-m_{t-1})\frac{\delta_{t|t-1}}{\delta_{t}}\sqrt{\bar{\alpha}}_{t-1}x_0 + \Bigl(m_{t-1}\delta_t - \frac{m_t(1-m_t)}{1-m_{t-1}}\alpha_t\delta_{t-1}\Bigr)\frac{\sqrt{\bar{\alpha}_{t-1}}}{\delta_t}y,\tilde{\delta}_tI\Bigr),
  \label{eq:new diffusion process t-1, t}
\end{dmath}
where $\tilde{\delta}_t$, the variance term of $q_{\text{cdiff}}(x_{t-1}|x_t,x_0,y)$, is
\begin{equation}
    \tilde{\delta}_t = \frac{\delta_{t|t-1} * \delta_t}{\delta_{t-1}}.
  \label{eq:new variance diffusion process t-1, t}
\end{equation}
To optimize the KL divergence term in Eq.~\eqref{eq:ELBO KL}, $\tilde{\delta}_t$ is also used as the variance of $p_{\text{cdiff}}(x_{t-1}|x_t,y)$ in Eq.~\eqref{eq:new reverse} to match $q_{\text{cdiff}}(x_{t-1}|x_t,x_0,y)$, and the coefficients $c_{xt}$, $c_{yt}$, $c_{\epsilon t}$ in Eq.~\eqref{eq:new reverse mean} are then be derived as:
\begin{align}
    c_{xt} &= \frac{1-m_t}{1-m_{t-1}}\frac{\delta_{t-1}}{\delta_{t}}\sqrt{\alpha}_t + (1-m_{t-1})\frac{\delta_{t|t-1}}{\delta_{t}}\frac{1}{\sqrt{\alpha}_t}, \label{eq:c_{xt}} \\
    c_{yt} &= (m_{t-1}\delta_t - \frac{m_t(1-m_t)}{1-m_{t-1}}\alpha_t\delta_{t-1})\frac{\sqrt{\bar{\alpha}_{t-1}}}{\delta_t}, \label{eq:c_{yt}} \\
    c_{\epsilon t} &= (1-m_{t-1})\frac{\delta_{t|t-1}}{\delta_t}\frac{\sqrt{1-\bar{\alpha}_t}}{\sqrt{\alpha_t}}.
  \label{eq:c_epsilon}
\end{align}

Now, given the explicit form of all distributions in Eq.~\eqref{eq:ELBO KL}, the ELBO to be optimized simplifies to
\begin{equation}
    c^\prime + \sum_{t=1}^{T}\kappa^\prime_t\mathbb{E}_{x_0,\epsilon,y} \parallel (\frac{m_t\sqrt{\bar{\alpha}_t}}{\sqrt{1-\bar{\alpha}_t}}{(y-x_0)} + \frac{\sqrt{\delta_t}}{\sqrt{1-\bar{\alpha}_t}}\epsilon) - \epsilon_\theta(x_t,y,t) \parallel^2_2
  \label{eq:ELBO_optimized}
\end{equation}
with constants $c^\prime$ and $\kappa^\prime_t$, and the $\epsilon$ is the Gaussian noise in $x_t$. Because we have the interpolation form of $x_t$ with the coefficient $m_t$ in Eq. \eqref{eq:new diffusion process t}, the optimization target in Eq. \eqref{eq:ELBO_optimized} keeps the simple form in training. Comparing to Eq. \eqref{eq:ELBO_optimized_original}, the $\epsilon_\theta(x_t,y,t)$ in the conditional diffusion model estimates both Gaussian noise $\epsilon$ and non-Gaussian noise $y-x_0$ in $x_t$. Therefore, the proportion of $y-x_0$ and $\epsilon$ coefficients is the same as y and the standard deviation in Eq. \eqref{eq:new diffusion process t}. 


\begin{algorithm}[tp]
  \caption{Training}\label{alg:training}
  \begin{algorithmic}
      \For{$i=1,2,\cdots,N_{\text{iter}}$}
        \State {Sample $(x_0,y) {\sim}q_{\text{data}}, \epsilon{\sim}\mathcal{N}(0,I),$ and} 
        \State {$t{\sim}\text{Uniform}(\{1,\cdots,T\})$}
        \State {$x_t = ((1-m_t)\sqrt{\bar{\alpha}_t}x_0 + m_t\sqrt{\bar{\alpha}_t}y) + \sqrt{\delta_t}\epsilon$ } 
        \State {Take gradient step on}
        \State {$\nabla_\theta \parallel \frac{1}{\sqrt{1-\bar{\alpha}_t}}(m_t\sqrt{\bar{\alpha}_t}{(y-x_0)} + \sqrt{\delta_t}\epsilon) - \epsilon_\theta(x_t,y,t) \parallel^2_2$}
        \State {according to Eq.~\eqref{eq:ELBO_optimized}}
      \EndFor
  \end{algorithmic}
\end{algorithm}

\subsection{CDiffuSE Training and Sampling Algorithm}
\label{sec:training}
In the conditional reverse process, according to Eq. \eqref{eq:new reverse} and \eqref{eq:ELBO_optimized}, $\epsilon_\theta(x_t,y,t)$ computes the combined noise, which is then deducted from  the combination of $x_t$ and $y$ 
to obtain cleaned data $x_{t-1}$. Finally, iterative application of this process over all $ T $ steps yields the clean signal $x_0$. The overall diffusion and reverse process of the conditional diffusion probabilistic models are described in Algorithms \ref{alg:training} and \ref{alg:sampling}. 
When the interpolation weight $ m_t $ of the real noise 
is set to $0$, the optimization target in Eq.~\eqref{eq:ELBO_optimized} and the reverse process in \eqref{eq:new reverse} becomes \eqref{eq:ELBO_optimized_original} and \eqref{eq:reverse_solution_original} as in the original diffusion probabilistic models.


%
%
%
 
In our previous study \cite{lu2021study}, a supportive reverse process was proposed as a less theoretically rigorous implementation to carry out the reverse process from the noisy speech (rather than isotropic Gaussian noise in the original reverse process) without changing the diffusion process. In our proposed CDiffuSE model, we remove the assumption that the real noise in $y$ follows the Gaussian distribution and avoid the mismatch issue between the diffusion and reverse process.

\begin{algorithm}[tp]
  \caption{Sampling}\label{alg:sampling}
  \begin{algorithmic}
      \State {Sample $x_T{\sim} \mathcal{N}(x_T,\sqrt{\bar{\alpha}_T}y,\delta_TI),$} 
      \For{$t=T,T-1,\cdots,1$}
        \State {Compute $c_{xt},c_{yt}$ and $c_{\epsilon t}$ using Eqs.~\eqref{eq:c_{xt}}, \eqref{eq:c_{yt}}, and \eqref{eq:c_epsilon}}
        \State {Sample $x_{t-1}\sim p_{\text{cdiff}}(x_{t-1}|x_t,y)=$}
        \State {$\mathcal{N}(x_{t-1};c_{xt}x_{t} + c_{yt} y - c_{\epsilon t} \epsilon_\theta(x_t,y,t), \tilde{\delta}_tI $}
        \}
      \EndFor
      \State \textbf{return} $x_0$
  \end{algorithmic}
\end{algorithm}

\section{Experiments}
In this section, we evaluate the performance of our approach against other generative speech enhancement models and we show generalization capabilities under conditions where state of the art approaches such as Demucs~\cite{defossez2020real} collapse. The samples of the CDiffuSE-enhanced signals can be found online\footnote{\label{note1}https://github.com/neillu23/CDiffuSE}.


\subsection{Experimental Setup}
\textbf{\textit{Dataset}}:
we evaluate the CDiffuSE model on the VoiceBank-DEMAND dataset \cite{valentini2016investigating}.
The dataset consists of 30 speakers from the VoiceBank corpus \cite{veaux2013voice}, which is further divided into a training set and a testing set with 28 and 2 speakers, respectively.
The training utterances are artificially contaminated with eight real-recorded noise samples from the DEMAND database \cite{thiemann2013diverse} and two artificially generated noise samples (babble and speech shaped) at 0, 5, 10, and 15 dB SNR levels, amounting to 11,572 utterances.
The testing utterances are mixed with different noise samples at 2.5, 7.5, 12.5, and 17.5 dB SNR levels, amounting to 824 utterances in total. We consider perceptual evaluation of speech quality (PESQ) \cite{rix2001perceptual},
prediction of the signal distortion (CSIG), prediction of the background intrusiveness (CBAK), and prediction of the overall speech quality (COVL) \cite{hu2007evaluation} as the evaluation metrics.
Higher scores mean better performance for all the metrics. 

\noindent \textbf{\textit{Model Architecture and Training}}:
we implement CDiffuSE based on the same model architecture and the same pre-training strategy with clean Mel-filterbank conditioner as that of DiffuSE reported in \cite{lu2021study}. We investigate two systems, namely Base and Large CDiffuSE, which respectively take $50$ and $200$ diffusion steps. The linearly spaced training noise schedule is reduced to $\beta_t \in [1\times10^{-4},0.035]$ for Base CDiffuSE, and to $\beta_t \in  [1\times10^{-4}, 0.0095]$ for Large CDiffuSE.
The interpolation parameter $m_t$ in Section~\ref{Sec:CDiffuSE diffusion} is set to $m_t = \sqrt{(1- \bar{\alpha}_t)/\sqrt{\bar{\alpha}_{t}}}$ which satisfies the $m_0 = 0$ and $m_t \approx 1$ requirement. 
We train both Base and Large CDiffuSE models for 300,000 iterations, based on an early stopping scheme. The batch size is set to 16 for Base CDiffuSE and to 15 for Large CDiffuSE. The fast sampling scheme \cite{kong2020diffwave} is used in the reverse processes with the inference schedule $\gamma_t =$ $ [0.0001, 0.001, 0.01, 0.05, 0.2, 0.35]$ for both Base CDiffuSE and Large CDiffuSE. 
The proposed CDiffuSE model performs enhancement in the time domain. After the reverse process is completed, the enhanced waveform further combine the original noisy signal with the ratio $0.2$ to recover the high frequency speech in the final enhanced waveform, as suggested in \cite{defossez2020real,abd2008speech}.

\begin{table}[b!]
\footnotesize
\centering
\caption{Results of DiffuSE and CDiffuSE on VoiceBank.}
\label{tab:ablation}
\begin{tabularx}{0.48\textwidth}{lXrrrr}
    \toprule
    Method & & PESQ & CSIG & CBAK & COVL \\
    \midrule
    \multicolumn{2}{l}{Unprocessed} & 1.97 & 3.35 & 2.44 & 2.63 \\
    \midrule
    DiffuSE (Base) \cite{lu2021study}  & & 2.41 & 3.61 & 2.81 & 2.99 \\
    CDiffuSE (Base) & & \textbf{2.44} & \textbf{3.66} & \textbf{2.83} & \textbf{3.03} \\
    \midrule
    DiffuSE (Large) \cite{lu2021study} & & 2.43 & 3.63 & 2.81 & 3.01 \\
    CDiffuSE (Large) & & \textbf{2.52} & \textbf{3.72} & \textbf{2.91} & \textbf{3.10} \\
    \bottomrule
\end{tabularx}\vspace{-0.5cm}
\end{table}

\begin{table}[tbp!]
\footnotesize
\centering
\caption{Performance comparison of CDiffuSE and time-domain generative models on VoiceBank. 
}
\label{tab:DiffuSE results}
\begin{tabularx}{0.48\textwidth}{Xrrrr}
    \toprule
    Method & PESQ & CSIG & CBAK & COVL \\
    \midrule	
    Unprocessed & 1.97 & 3.35 & 2.44 & 2.63 \\
    SEGAN \cite{pascual2017segan}& 2.16 & 3.48 & 2.94 & 2.80 \\
    DSEGAN \cite{phan2020improving}& 2.39 & 3.46 & \textbf{3.11} & 2.90 \\
    SE-Flow \cite{strauss2021flow}& 2.28 & 3.70 & 3.03 & 2.97 \\
    \midrule	
    CDiffuSE (Base) & \textbf{2.44} & 3.66  & 2.83  & \textbf{3.03}\\
    CDiffuSE (Large) & \textbf{2.52} & \textbf{3.72}  & 2.91 & \textbf{3.10}\\
    \bottomrule
\end{tabularx}\vspace{-0.5cm}
\end{table}

\begin{table}[tbp!]
\footnotesize
\centering
\caption{Comparison of CDiffuSE and discriminative models. 
}
\label{tab:generalization}
\subfigure[Trained and tested on VoiceBank (\textbf{matched} condition).]{
\footnotesize
\label{tab:generalization_vb}
\begin{tabularx}{0.48\textwidth}{Xrrrr}
    \toprule
    Method & PESQ & CSIG & CBAK & COVL \\
    \midrule
    Unprocessed & 1.97 & 3.35 & 2.44 & 2.63 \\
    \midrule
    WaveCRN \cite{hsieh2020wavecrn}& 2.63 & 3.95 & 3.06 & 3.29 \\
    Demucs \cite{defossez2020real} & 2.65 & \textbf{3.99} & \textbf{3.33} & \textbf{3.32 }\\
    Conv-TasNet \cite{luo2019conv}& \textbf{2.84} & 2.33 & 2.62 & 2.51 \\
    \midrule
    CDiffuSE (Large) & 2.52 & 3.72 & 2.91 & 3.10 \\
    \bottomrule
\end{tabularx}
}
\vspace{-0.5cm}
\subfigure[Trained on VoiceBank, tested on CHiME-4 (\textbf{mismatched} condition).]{
\footnotesize
\label{tab:generalization_chime}
\begin{tabularx}{0.48\textwidth}{Xrrrr}
    \toprule
    Method & PESQ & CSIG & CBAK & COVL \\
    \midrule
    Unprocessed & 1.27 & 2.61 & 1.93 & 1.88 \\
    \midrule
    WaveCRN \cite{hsieh2020wavecrn}& 1.43 & 2.53 & 2.03 & 1.91 \\
    Demucs \cite{defossez2020real} & 1.38 & 2.50 & 2.08 & 1.88 \\
    Conv-TasNet \cite{luo2019conv}& 1.63 &	1.70 & 1.82 & 1.54  \\
    \midrule
    CDiffuSE (Large) & \textbf{1.66}  & \textbf{2.98} & \textbf{2.19} & \textbf{2.27}\\
    \bottomrule
\end{tabularx}
}
\end{table}

\subsection{Evaluation results}



\subsubsection{Results on VoiceBank-DEMAND}

In Table \ref{tab:ablation}, we report the results of CDiffuSE and DiffuSE using the supportive reverse process from~\cite{lu2021study}. 
As expected, the large models for DiffuSE and CDiffuSE both outperform the smaller base models.
Moreover, CDiffuSE shows improved performance over the diffusion probabilistic model baseline DiffuSE.
Note that the key to success here is that CDiffuSE has had direct access to the noisy data while learning the reverse diffusion process, allowing it to actively compensate for the noise characteristics in the input signals.
Being able to leverage noise from the input signal, our approach improves on all the metrics, confirming that the theoretically sound CDiffuSE leads to improved results in practice.
%
We additionally compare CDiffuSE to other time-domain generative models, namely SEGAN~\cite{pascual2017segan}, SE-Flow~\cite{strauss2021flow}, and improved deep SEGAN (DSEGAN)~\cite{phan2020improving}.
CDiffuSE outperforms its competitors on all metrics - with the exception of CBAK - and achieves a particularly significant improvement in PESQ, see Table~\ref{tab:DiffuSE results}. 

\vspace{-.2cm}
\subsubsection{Results on CHiME-4}

Generative models typically aim to fit the distribution of the training samples instead of optimizing a point-wise $ L_p $-loss.
This property has made them state of the art in applications like text-to-speech and vocoding~\cite{oord2016wavenet, kong2020hifi} and also makes them more robust against domain shifts in the input data.

In this section, we investigate this property of our proposed CDiffuSE.
We compare the generalization abilities of our approach to other, $ L_p $-loss based approaches and demonstrate that our approach is particularly resistant towards shifts in noise characteristics of the speech data.
The models in this section are trained on VoiceBank-DEMAND and evaluated on the simulated test data of CHiME-4~\cite{vincent2017analysis}.
The CHiME-4 simulated test data is created based on real-recorded noises from four real-world environments (including street, pedestrian areas, cafeteria and bus) based on four speakers.
We use the signals from the fifth microphone for evaluation.

As mentioned previously and as Table~\ref{tab:generalization_vb} shows, generative speech enhancement models are still lagging behind the performance of their regressive counterparts.
A model from the latter category trained on VoiceBank and evaluated on the VoiceBank test set performs far better than most generative methods.
Particularly, Demucs~\cite{defossez2020real} and Conv-TasNet~\cite{luo2019conv} outperform our CDiffuSE, which was the strongest generative model in Table~\ref{tab:DiffuSE results}.

Given a domain shift in test data, however, regression based approaches such as Demucs, Conv-TasNet, and WaveCRN suffer from a significant drop in performance, see Table~\ref{tab:generalization_chime}.
Different signal characteristics between the VoiceBank training data and the CHiME-4 test set suffice to let the evaluation scores fall drastically, in some cases even below the scores of unprocessed data.
Our proposed CDiffuSE, on the contrary, proves to be much more resilient against such shifts in signal characteristics.
While the scores on the CHiME-4 test set are lower than the VoiceBank scores, CDiffuSE degrades to a much smaller degree than its regressive competitors, leaving it with the best scores on the CHiME-4 test data and demonstrating its high robustness to variation in noise characteristics.

\vspace{-.3cm}
\section{Conclusion}
\vspace{-.2cm}

We proposed CDiffuSE, a conditional diffusion probabilistic model that can explore noise characteristics from the noisy input signal explicitly and thereby adapts better to non-Gaussian noise statistics in real-world speech enhancement problems.
We showed that our model is a strict generalization of the original diffusion probabilistic model and achieves state of the art results compared to other generative speech enhancement approaches.
In contrast to non-generative approaches, our method exposes great generalization capabilities to speech data with noise characteristics not observed in the training data.
We were able to show that CDiffuSE maintains strong performance when regression-based approaches such as Demucs and Conv-TasNet collapse.

\label{sec:typestyle}
\begin{spacing}{0.85}
\bibliographystyle{IEEEtran}
\bibliography{refs}

\begin{thebibliography}{10}
\providecommand{\url}[1]{#1}
\csname url@samestyle\endcsname
\providecommand{\newblock}{\relax}
\providecommand{\bibinfo}[2]{#2}
\providecommand{\BIBentrySTDinterwordspacing}{\spaceskip=0pt\relax}
\providecommand{\BIBentryALTinterwordstretchfactor}{4}
\providecommand{\BIBentryALTinterwordspacing}{\spaceskip=\fontdimen2\font plus
\BIBentryALTinterwordstretchfactor\fontdimen3\font minus
  \fontdimen4\font\relax}
\providecommand{\BIBforeignlanguage}[2]{{%
\expandafter\ifx\csname l@#1\endcsname\relax
\typeout{** WARNING: IEEEtran.bst: No hyphenation pattern has been}%
\typeout{** loaded for the language `#1'. Using the pattern for}%
\typeout{** the default language instead.}%
\else
\language=\csname l@#1\endcsname
\fi
#2}}
\providecommand{\BIBdecl}{\relax}
\BIBdecl

\bibitem{li2014overview}
J.~Li, L.~Deng, Y.~Gong, and R.~Haeb-Umbach, ``An overview of noise-robust
  automatic speech recognition,'' \emph{IEEE/ACM Transactions on Audio, Speech,
  and Language Processing}, vol.~22, no.~4, pp. 745--777, 2014.

\bibitem{HaebUmbach2020Overview}
R.~Haeb-Umbach, J.~Heymann, L.~Drude, S.~Watanabe, M.~Delcroix, and
  T.~Nakatani, ``Far-field automatic speech recognition,'' \emph{Proceedings of
  the IEEE}, 2020.

\bibitem{healy2019optimal}
E.~W. Healy, J.~L. Vasko, and D.~Wang, ``The optimal threshold for removing
  noise from speech is similar across normal and impaired hearing—a
  time-frequency masking study,'' \emph{The Journal of the Acoustical Society
  of America}, vol. 145, no.~6, pp. EL581--EL586, 2019.

\bibitem{H.L.Hansen2015}
J.~{H.L. Hansen} and T.~Hasan, ``Speaker recognition by machines and humans: A
  tutorial review,'' \emph{IEEE Signal Processing Magazine}, vol.~32, no.~6,
  pp. 74--99, 2015.

\bibitem{michelsanti2017conditional}
D.~Michelsanti and Z.-H. Tan, ``Conditional generative adversarial networks for
  speech enhancement and noise-robust speaker verification,'' \emph{arXiv
  preprint arXiv:1709.01703}, 2017.

\bibitem{wang2018supervised}
D.~Wang and J.~Chen, ``Supervised speech separation based on deep learning: An
  overview,'' \emph{IEEE/ACM Transactions on Audio, Speech, and Language
  Processing}, vol.~26, no.~10, pp. 1702--1726, 2018.

\bibitem{lu2013speech}
X.~Lu, Y.~Tsao, S.~Matsuda, and C.~Hori, ``Speech enhancement based on deep
  denoising autoencoder,'' in \emph{Proc. Interspeech 2013}.

\bibitem{fu2018end}
S.-W. Fu, T.-W. Wang, Y.~Tsao, X.~Lu, and H.~Kawai, ``End-to-end waveform
  utterance enhancement for direct evaluation metrics optimization by fully
  convolutional neural networks,'' \emph{IEEE/ACM Transactions on Audio,
  Speech, and Language Processing}, vol.~26, no.~9, pp. 1570--1584, 2018.

\bibitem{koizumi2018dnn}
Y.~Koizumi, K.~Niwa, Y.~Hioka, K.~Kobayashi, and Y.~Haneda, ``{DNN}-based
  source enhancement to increase objective sound quality assessment score,''
  \emph{IEEE/ACM Transactions on Audio, Speech, and Language Processing},
  vol.~26, no.~10, pp. 1780--1792, 2018.

\bibitem{fu2019metricgan}
S.-W. Fu, C.-F. Liao, Y.~Tsao, and S.-D. Lin, ``{MetricGAN}: Generative
  adversarial networks based black-box metric scores optimization for speech
  enhancement,'' in \emph{Proc. ICML 2019}.

\bibitem{pascual2017segan}
S.~Pascual, A.~Bonafonte, and J.~Serra, ``{SEGAN}: Speech enhancement
  generative adversarial network,'' \emph{arXiv preprint arXiv:1703.09452},
  2017.

\bibitem{soni2018time}
M.~H. Soni, N.~Shah, and H.~A. Patil, ``Time-frequency masking-based speech
  enhancement using generative adversarial network,'' in \emph{Proc. ICASSP
  2018}.

\bibitem{qian2017speech}
K.~Qian, Y.~Zhang, S.~Chang, X.~Yang, D.~Flor{\^e}ncio, and
  M.~Hasegawa-Johnson, ``Speech enhancement using bayesian {Wavenet},'' in
  \emph{Proc. Interspeech 2017}.

\bibitem{leglaive2020recurrent}
S.~Leglaive, X.~Alameda-Pineda, L.~Girin, and R.~Horaud, ``A recurrent
  variational autoencoder for speech enhancement,'' in \emph{Proc. ICASSP
  2020}.

\bibitem{strauss2021flow}
M.~Strauss and B.~Edler, ``A flow-based neural network for time domain speech
  enhancement,'' in \emph{Proc. ICASSP 2021}.

\bibitem{defossez2020real}
A.~Defossez, G.~Synnaeve, and Y.~Adi, ``Real time speech enhancement in the
  waveform domain,'' \emph{arXiv preprint arXiv:2006.12847}, 2020.

\bibitem{wang2019bridging}
P.~Wang, K.~Tan \emph{et~al.}, ``Bridging the gap between monaural speech
  enhancement and recognition with distortion-independent acoustic modeling,''
  \emph{IEEE/ACM Transactions on Audio, Speech, and Language Processing},
  vol.~28, pp. 39--48, 2019.

\bibitem{bagchi2018spectral}
D.~Bagchi, P.~Plantinga, A.~Stiff, and E.~Fosler-Lussier, ``Spectral feature
  mapping with mimic loss for robust speech recognition,'' in \emph{Proc.
  ICASSP 2018}.

\bibitem{gao2016snr}
T.~Gao, J.~Du, L.-R. Dai, and C.-H. Lee, ``Snr-based progressive learning of
  deep neural network for speech enhancement.'' in \emph{Proc. Interspeech
  2016}.

\bibitem{sohl2015deep}
J.~Sohl-Dickstein, E.~Weiss, N.~Maheswaranathan, and S.~Ganguli, ``Deep
  unsupervised learning using nonequilibrium thermodynamics,'' in \emph{Proc.
  ICML 2015}.

\bibitem{ho2020denoising}
J.~Ho, A.~Jain, and P.~Abbeel, ``Denoising diffusion probabilistic models,''
  \emph{arXiv preprint arXiv:2006.11239}, 2020.

\bibitem{nichol2021improved}
A.~Nichol and P.~Dhariwal, ``Improved denoising diffusion probabilistic
  models,'' \emph{arXiv preprint arXiv:2102.09672}, 2021.

\bibitem{kong2020diffwave}
Z.~Kong, W.~Ping, J.~Huang, K.~Zhao, and B.~Catanzaro, ``Diffwave: A versatile
  diffusion model for audio synthesis,'' \emph{arXiv preprint
  arXiv:2009.09761}, 2020.

\bibitem{liu2021diffsvc}
S.~Liu, Y.~Cao, D.~Su, and H.~Meng, ``Diffsvc: A diffusion probabilistic model
  for singing voice conversion,'' \emph{arXiv preprint arXiv:2105.13871}, 2021.

\bibitem{lu2021study}
Y.-J. Lu, Y.~Tsao, and S.~Watanabe, ``A study on speech enhancement based on
  diffusion probabilistic model,'' \emph{arXiv preprint arXiv:2107.11876},
  2021.

\bibitem{valentini2016investigating}
C.~Valentini-Botinhao, X.~Wang, S.~Takaki, and J.~Yamagishi, ``Investigating
  {RNN}-based speech enhancement methods for noise-robust text-to-speech.'' in
  \emph{SSW}, 2016, pp. 146--152.

\bibitem{veaux2013voice}
C.~Veaux, J.~Yamagishi, and S.~King, ``The voice bank corpus: Design,
  collection and data analysis of a large regional accent speech database,'' in
  \emph{Proc. CASLRE 2013}.

\bibitem{thiemann2013diverse}
J.~Thiemann, N.~Ito, and E.~Vincent, ``The diverse environments multi-channel
  acoustic noise database (demand): A database of multichannel environmental
  noise recordings,'' in \emph{Proceedings of Meetings on Acoustics}, vol.~19,
  no.~1, 2013, p. 035081.

\bibitem{rix2001perceptual}
A.~W. Rix, J.~G. Beerends, M.~P. Hollier, and A.~P. Hekstra, ``Perceptual
  evaluation of speech quality (pesq)-a new method for speech quality
  assessment of telephone networks and codecs,'' vol.~2, 2001, pp. 749--752.

\bibitem{hu2007evaluation}
Y.~Hu and P.~C. Loizou, ``Evaluation of objective quality measures for speech
  enhancement,'' \emph{IEEE Transactions on audio, speech, and language
  processing}, vol.~16, no.~1, pp. 229--238, 2007.

\bibitem{abd2008speech}
M.~Abd El-Fattah, M.~I. Dessouky, S.~Diab, and F.~Abd El-Samie, ``Speech
  enhancement using an adaptive wiener filtering approach,'' \emph{Progress In
  Electromagnetics Research M}, vol.~4, pp. 167--184, 2008.

\bibitem{phan2020improving}
H.~Phan, I.~V. McLoughlin, L.~Pham, O.~Y. Ch{\'e}n, P.~Koch, M.~De~Vos, and
  A.~Mertins, ``Improving gans for speech enhancement,'' \emph{IEEE Signal
  Processing Letters}, vol.~27, pp. 1700--1704, 2020.

\bibitem{hsieh2020wavecrn}
T.-A. Hsieh, H.-M. Wang, X.~Lu, and Y.~Tsao, ``Wavecrn: An efficient
  convolutional recurrent neural network for end-to-end speech enhancement,''
  \emph{IEEE Signal Processing Letters}, vol.~27, pp. 2149--2153, 2020.

\bibitem{luo2019conv}
Y.~Luo and N.~Mesgarani, ``Conv-{TasNet}: Surpassing ideal time--frequency
  magnitude masking for speech separation,'' \emph{IEEE/ACM Transactions on
  Audio, Speech, and Language Processing}, vol.~27, no.~8, pp. 1256--1266,
  2019.

\bibitem{oord2016wavenet}
A.~V.~D. Oord, S.~Dieleman, H.~Zen, K.~Simonyan, O.~Vinyals, A.~Graves,
  N.~Kalchbrenner, A.~Senior, and K.~Kavukcuoglu, ``Wavenet: A generative model
  for raw audio,'' \emph{arXiv preprint arXiv:1609.03499}, 2016.

\bibitem{kong2020hifi}
J.~Kong, J.~Kim, and J.~Bae, ``Hifi-gan: Generative adversarial networks for
  efficient and high fidelity speech synthesis,'' \emph{arXiv preprint
  arXiv:2010.05646}, 2020.

\bibitem{vincent2017analysis}
E.~Vincent, S.~Watanabe, A.~A. Nugraha, J.~Barker, and R.~Marxer, ``An analysis
  of environment, microphone and data simulation mismatches in robust speech
  recognition,'' \emph{Computer Speech \& Language}, vol.~46, pp. 535--557,
  2017.

\end{thebibliography}
\end{spacing}
\end{document}